\documentclass{article}
\usepackage{spconf,amsmath,amssymb,graphicx,pgfplots,tikz}
\usepackage{hyperref}
\usepackage{caption}
\usetikzlibrary{math}

\title{CGCNN: COMPLEX GABOR CONVOLUTIONAL NEURAL NETWORK ON RAW SPEECH}
\name{Paul-Gauthier No\'e$^1$, Titouan Parcollet$^{1,2}$, Mohamed Morchid$^1$}
\address{$^1$LIA, Universit\'e d'Avignon, France\\$^2$University of Oxford, UK}
\begin{document}
%
\maketitle
\begin{abstract}

Convolutional Neural Networks (CNN) have been used in Automatic Speech Recognition (ASR) to learn representations directly from the raw signal instead of hand-crafted acoustic features, providing a richer and lossless input signal. Recent researches propose to inject prior acoustic knowledge to the first convolutional layer by integrating the shape of the impulse responses in order to increase both the interpretability of the learnt acoustic model, and its performances. We propose to combine the complex Gabor filter with complex-valued deep neural networks to replace usual CNN weights kernels, to fully take advantage of its optimal time-frequency resolution and of the complex domain. The conducted experiments on the TIMIT phoneme recognition task shows that the proposed approach reaches top-of-the-line performances while remaining interpretable.
\end{abstract}
\begin{keywords}
SincNet, complex neural networks, Gabor filters, speech recognition.
\end{keywords}
\section{Introduction}
\label{sec:intro}

The task of Automatic Speech Recognition (ASR) is far from being solved and represents an active research field \cite{baidu1015deepspeech,pytorch-kaldi,watanabe2018espnet,li2019jasper}. More precisely ASR systems are either hybrid DNN-HMM, with multiple sub-blocks trained separately \cite{Povey_ASRU2011,mirco2017timit}, or End-to-End (E2E), with various Neural Networks (NN) learnt accordingly to a joint training procedure \cite{graves2014towards,zhang2017towards,kim2017joint}. Despite numerous architectures and training investigations, the input feature representation remains mostly unchanged with traditional handcrafted acoustic features such as Mel-filter-banks.

Nonetheless, another field of speech recognition recently obtained promising performances while operating at the raw waveform level \cite{palaz2013end,tuske2014acoustic,hoshen2015speech,zeghidour2018end}. In these works, Convolutional Neural Networks (CNN) have been used directly on the raw speech signal to rapidly consume the large number of data points (\textit{e.g.} $16,000$ per second for an audio sampled at $16$kHz). Unfortunately, and as demonstrated in \cite{sincnet,ravanelli2018speech} traditional CNN kernel filters are not efficient at learning common acoustic features due to the lack of constraint on neural parameters. To alleviate the latter issue, the authors proposed a novel convolutional layer named SincNet incorporating prior speech processing knowledge in an efficient and interpretable operation. More precisely, SincNet filters are initialized and constrained following specific acoustic filters to produce easily interpretable filtering of the input waveform \cite{sincnet}. As a matter of fact, processing the raw waveform theoretically has numerous advantages including a higher richness of the input signal alongside with a lossless natural representation of the features. Recently, other filters have been proposed for SincNet to further increase its performances \cite{kbfilters}. In the latter, Gaussian filters have been suggested without motivations on their major properties in terms of time and frequency localization \cite{gabor}. Furthermore, the authors only proposed to consider the real-part to feed the standard SincNet architecture. We propose to extend this work to the complete complex Gabor filters to take advantage of both the complex-valued representation and a better time-frequency resolution due to the Gaussian shape. Other recent works have used Gabor filters instead of the triangular ones to operate over FBANK acoustic features \cite{g-bank}. In this case, cutoff frequencies of the filters are fixed on the Mel-scale and require fine tuning. Gabor filters have also been used to initialize end-to-end filterbanks learning \cite{learningfbank}. In the latter works, the filter frequency positions are fixed. We propose to learn all these parameters jointly to the rest of the neural network model to allow the Gabor filters to perfectly match the considered task. 

Nonetheless, complex Gabor filters produce a complex-valued filtered signal that must be processed with a dedicated complex-valued model to further fully exploit this input representation. Fortunately, Complex-Valued Convolutional Neural Networks (CVCNN) have been recently introduced with applications to image and speech processing \cite{chineb, phaseaware}. Thus, we propose to enhance the original real-valued SincNet with the proposed complex Gabor filters, and to process the filtered signal with a CVCNN respectful of the complex algebra. Contributions of the paper are summarized as:

\begin{enumerate}
\item The optimal time and frequency localization compromise of Gabor filters is presented (Section \ref{sec:gabor}).
\item SincNet \cite{sincnet} is extended to the complex-valued space with a complex Gabor filter and CVCNNs. The model is also released for reproducibility \footnote{\label{git}\url{https://github.com/NOEPG/pytorch-kaldi}} (Section \ref{sec:ccnn}). 
\item Evaluate and analyse the proposed complex gabor-based SincNet on a phoneme recognition task with the TIMIT dataset (Section \ref{sec:exps}). 
\end{enumerate}

The conducted experiments show that our approach reaches promising performances with a natural and efficient filtered representation of the raw waveform, while retaining the interpretability property of the original SincNet.

\section{UNCERTAINTY PRINCIPLE AND GABOR FILTERING}
\label{sec:gabor}

\begin{figure}[!t]
    \centering
    \includegraphics[width=60mm]{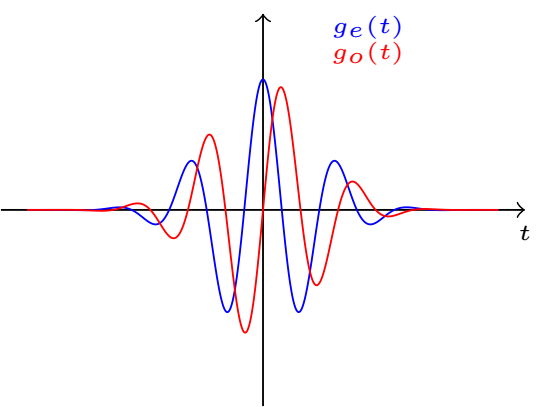}
    \caption{Illustration of the real ($g_e$) and imaginary ($g_o$)  parts of the complex Gabor filter impulse response.}
    \label{fig:imp_rep}
\end{figure}

This section motivates the use of Gabor filters to replace the standard sinc approach of SincNet \cite{sincnet}.  

First, it is important to notice that time-frequency resolution is limited by the uncertainty principle. More precisely, the product of the time and frequency spread of a filter is superior to a constant. Given any function $y \in L^{2}(\mathbb{R})$ and its Fourier transform $\hat{y}$, their time $e_{y}$ and frequency $e_{\hat{y}}$ localizations are respectively:

\begin{equation}
\begin{split}
  e_{y} &=\frac{1}{||y||^{2}} \int_{-\infty}^{+\infty}t|y(t)|^{2}dt.\\
  e_{\hat{y}} &=\frac{1}{||\hat{y}||^{2}} \int_{-\infty}^{+\infty}f|\hat{y}(f)|^{2}df.
\end{split}
\end{equation}

Then, their time and frequency spreads are defined by their variance $v_y$ and $v_{\hat{y}}$:

\begin{equation}
\begin{split}
  v_{y} &=\frac{1}{||y||^{2}} \int_{-\infty}^{+\infty}(t-e_{y})^{2}|y(t)|^{2}dt.\\
  v_{\hat{y}} &=\frac{1}{||\hat{y}||^{2}} \int_{-\infty}^{+\infty}(f-e_{\hat{y}})^{2}|\hat{y}(f)|^{2}df.
\end{split}
\end{equation}

The uncertainty principle states that $v_{y}v_{\hat{y}} \geq \frac{1}{16\pi^{2}}$ \cite{mallat}. Thus a function can not have an infinite narrow localization both in time and frequency domain. However, this inequality becomes an equality for Gaussian shape functions such as Gabor filter \cite{mallat} and thus provides the best time-frequency resolution compromise.

In particular, the complex impulse response of the Gabor filter is defined as follows:

\begin{equation}
\begin{split}
  g(t) & = w_{\sigma}(t)e^{i2\pi f_{0}t},\\
       & =g_{e}(t)+ig_{o}(t),\\
  w_{\sigma}(t)&=\frac{1}{\sqrt{2\pi}\sigma}e^{-\frac{t^2}{2\sigma^2}},
\end{split}
\end{equation}
with $\sigma$ the standard deviation of the temporal Gaussian window $w_{\sigma}(t)$, $f_{0}$ the center frequency of the filter and $g_{e}(t)$ and $g_{o}(t)$ the real and imaginary part of the impulse response respectively (Figure \ref{fig:imp_rep}). 

The Gabor filter has the following Gaussian frequency response:\\

\begin{equation}
  G(f)=e^{-2\pi^{2}\sigma^{2}(f-f_{0})^{2}}.
\end{equation}

\begin{figure}[!t]
    \centering
    \includegraphics[width=60mm]{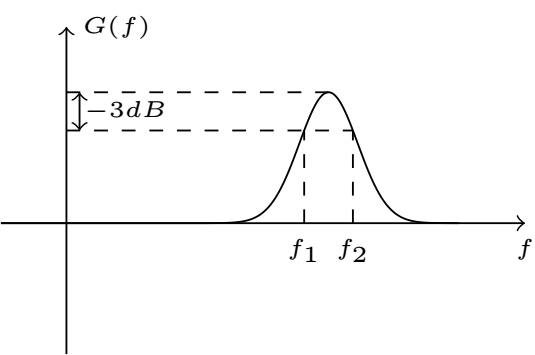}
    \caption{Illustration of the frequency response of the complex Gabor filter with $f_{1}$ and $f_{2}$ respectively the low and high cutoff frequencies.}
    \label{fig:freq_rep}
\end{figure}

For the sake of interpretability, it is feasible to express $\sigma$ and $f_{0}$ in terms of the cutoff frequencies:

\begin{equation}
  \sigma = \frac{A}{\pi(f_{2}-f_{1})},\thickspace
  f_{0} = \frac{f_{1}+f_{2}}{2},
\end{equation}
with $f_{1}$ and $f_{2}$  the low and high ($-3dB$) cutoff frequencies respectively, and $A=\sqrt{\frac{3ln(10)}{10}}$ a constant. The Gabor frequency response is depicted in Figure.\ref{fig:freq_rep} \\

\section{CONVOLUTIONAL NEURAL NETWORKS ON THE RAW WAVEFORM}
\label{sec:ccnn}

This section details the proposed complex-valued model to directly operate of the raw speech waveform. First, SincNet is presented (Section \ref{subsec:sincnet}). Then, we introduce the complex Gabor filtering alongside with the complex-valued architecture (Section \ref{subsec:gabor}). 

\subsection{SincNet}
\label{subsec:sincnet}

Traditional parametric CNNs operate over the raw waveform by performing multiple time-domain convolutions between the input signal and a certain finite impulse response. Therefore, SincNet \cite{sincnet} proposes to use a sinc function to obtain the impulse response $h(t)$ at time $t$ of a CNN as:

\begin{equation}
\label{eq:sinc}
    h(t)=2f_{2}sinc(2\pi f_{2} t) - 2f_{1}sinc(2\pi f_{1} t),
\end{equation}
with sinc($x$)=sin($x$)$/x$, $f_{1}$ and $f_{2}$ the two cutoff frequencies and $f_{2} \geq f_{1}$. Such filter has a rectangular frequency response described as:

\begin{equation}
    H(f) = rect_{2f_{2}}(f)-rect_{2f_{1}}(f),
\end{equation}
with $rect_{w}(.)$ the rectangular function of width $w$ centered in $0$. Finally, cutoff frequencies are the two only trainable parameters reducing drastically the number of neural parameters compared to CNNs. Then, multiple CNN layers are stacked to reduce the signal dimension, before being fed into a classifier. 

\subsection{Complex Gabor Filtering with CVCNN}
\label{subsec:gabor}

We propose to replace the sinc based filter of Eq. \ref{eq:sinc} with a complex-valued Gabor one, for a better time-frequency resolution. Therefore, let $g(t)$ be the complex Gabor filter impulse response with $g_e(t)$ and $g_o(t)$ the corresponding real and imaginary parts. The convolution of an input signal $x(t)$ with $g(t)$ at time $t$ is written as:
\begin{equation}
  x(t)*g(t) = x(t)*g_{e}(t) + i (x(t)*g_{o}(t)).
\end{equation}

The obtained filtered signal is of the form $z = a+\textbf{i}b$ lying on the complex plane, and is an approximation of analytic signal (\textit{i.e.} without negative frequencies) \cite{clifford}. We propose to fully take this complex representation into consideration by further processing it with complex-valued neural networks layers only.

Therefore, the extracted features are then fed into multiple CVCNNs followed by complex layer-normalization to reduce the signal dimension. Finally, the reduced complex features are fed into a complex fully-connected classifier (CVDNN). In the same manner as SincNet \cite{pytorch-kaldi}, the two cutoff frequencies of the filters are the only parameters learnt in the first layer, thus reducing the number of neural parameters. 

\section{EXPERIMENTS}
\label{sec:exps}

The proposed approach is evaluated on the TIMIT phoneme recognition task \cite{timit}. The latter dataset is composed of a standard $462$-speaker training dataset, a $50$-speakers development dataset and a core test dataset of $192$ sentences. During the experiments, the dialect sentences of the training dataset are removed.  

\subsection{Models Achitecture}

Our model starts with four CVCNNs with respectively $128$, $60$, $60$ and $60$ complex filters of kernel size $129$, $5$, $5$, $3$ with Complex Gabor filters used at the first layer only. Then, a Complex-Valued Multilayer Perceptron (CMLP) composed of $5$ layers with $1024$ complex hidden units is added. The ReLU activation function \cite{nair2010rectified} is used accross all the layers. Complex layer-normalization is done on each layer of the CVCNN, while complex batch-normalization is applied on the CMLP layers. Maxpooling is applied on the convolutional part to further reduce the dimension of the signal. Finally the $softmax$ function is applied to obtain the posterior probability over the HMM states. Monophone regularization is also used to smooth the training \cite{monophone}.  

We propose to compare this approach with a real-valued equivalent baseline. Thus, the input layer has the same number of filters. Then, the dimensions of the other layers are increased to obtain a comparable number of neural parameters $20M$.

Both models are fed with $200$ms speech signal chunks with an overlap of $10$ms. They are trained using standard stochastic gradient descent with a batch size of $128$ for $20$ epochs. The learning rate is annealed with respect to a certain threshold on the validation loss evolution, alleviating the risk of overfitting and ensuring an optimal convergence for both models. A dropout probability of $0.15$ is also set for each layer except the classification ones.

Models and experiments are run within the Pytorch-Kaldi toolkit \cite{pytorch-kaldi}.

\subsection{Results and Discussions}

Models are evaluated by averaging the Phone Error Rate (PER) observed on the validation and test datasets for $5$ runs with the TIMIT phoneme recognition task \cite{timit}. Results are shown in Table \ref{table:results}.

First, it is worth underlying that the obtained results are comparable to state-of-the-art performances on the TIMIT task from the raw waveform. Then, while both real and complex-valued Gabor models obtain similar averaged performances, a best PER of $16.7\%$ is obtained with the complex alternative compared to $16.9\%$ for the real-valued one. In fact, both real and complex filters have the same frequency response in the positive domain. But using complex quadratic filters that produce analytic signal for which the complex Gabor filtered signal is an approximation could help for instantaneous frequency estimation \cite{clifford} and preserves the phase information that can be useful for other tasks such as speaker recognition.

\begin{table*}[t!]
\centering
\caption{Results obtained with different ASR systems on the TIMIT phoneme recognition tasks. Valid. denotes the validation dataset, and CTC the Connectionist Temporal Classification training scheme.  Results are expressed in Phoneme Error Rate (\textit{i.e.} lower is better).\\}

\scalebox{1}{
    \begin{tabular}{cccc}
        \hline\hline
        \textbf{Model} & \textbf{Valid.\%} & \textbf{Avg. Test\%}  & \textbf{Best Test\%} \\
        \hline
        Gabor-CNN-CTC \cite{g-bank} & - & 18.8  & 18.5 \\
        SincNet \cite{pytorch-kaldi} & - & 17.2 & - \\
        GaborReal & 15.2 & 17.2 & 16.9  \\
        GaborComplex & 15.2 & 17.1 & 16.7  \\
        \hline
    \end{tabular}
    \label{table:results}
}

\end{table*}

\begin{figure}[!t]
    \centering

    \begin{tikzpicture}
        \begin{axis}[
        axis x line=middle,
            axis y line=middle,
            enlarge y limits=true,
            width=8cm, height=6cm,     
            ylabel=$f_{2}$,
            xlabel=$f_{1}$,
		    legend style={at={(0.97,0.4)}},
		    ]
           
        \addplot [color = blue, opacity=0.65] [only marks,mark size=0.7pt] table {gaborC/freqs8000};
        \addlegendentry{\tiny{Complex Gabor}}      
      
        \addplot [color = red, opacity=0.65] [only marks,mark size=0.7pt] table {gaborR/freqs8000};
        \addlegendentry{\tiny{Real Gabor}}

        \end{axis}
    \end{tikzpicture}
    
    \caption{Illustration of the cutoff frequencies learnt with the Real and the Complex Gabor CNN.}
    \label{fig:learnt_freq}
\end{figure}
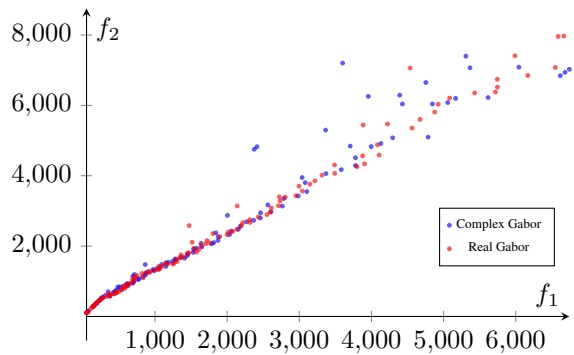

Then, Figure \ref{fig:learnt_freq} illustrates the learnt cutoff frequencies. As expected, distributions are similar along a straight line for both models. In fact, both models roughly learn the same couples of cutoff frequencies due to an equivalent frequency response shape. 

\begin{figure}[!t]
 \centering
    \includegraphics[width=8cm]{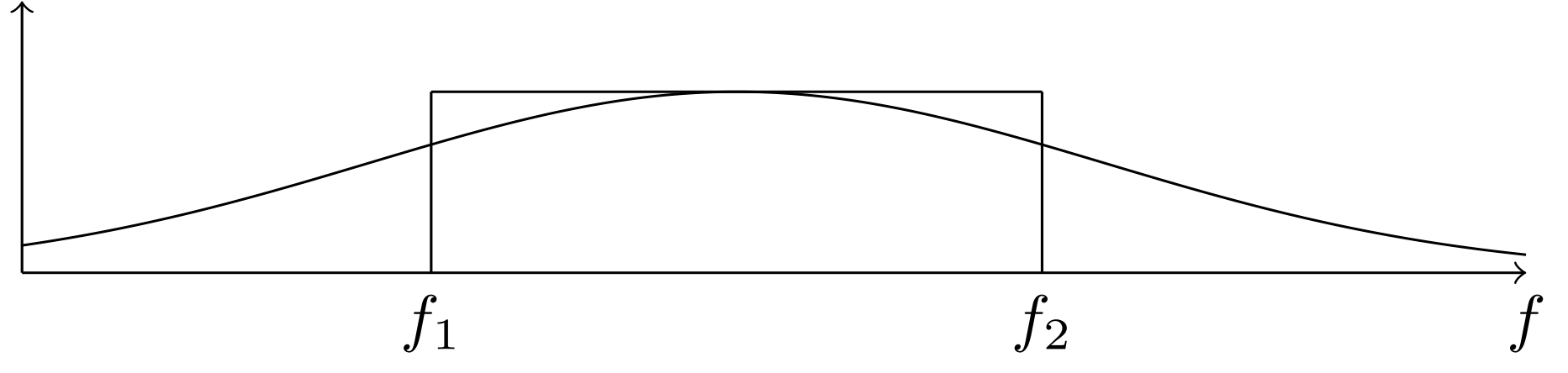}
    \includegraphics[width=8cm]{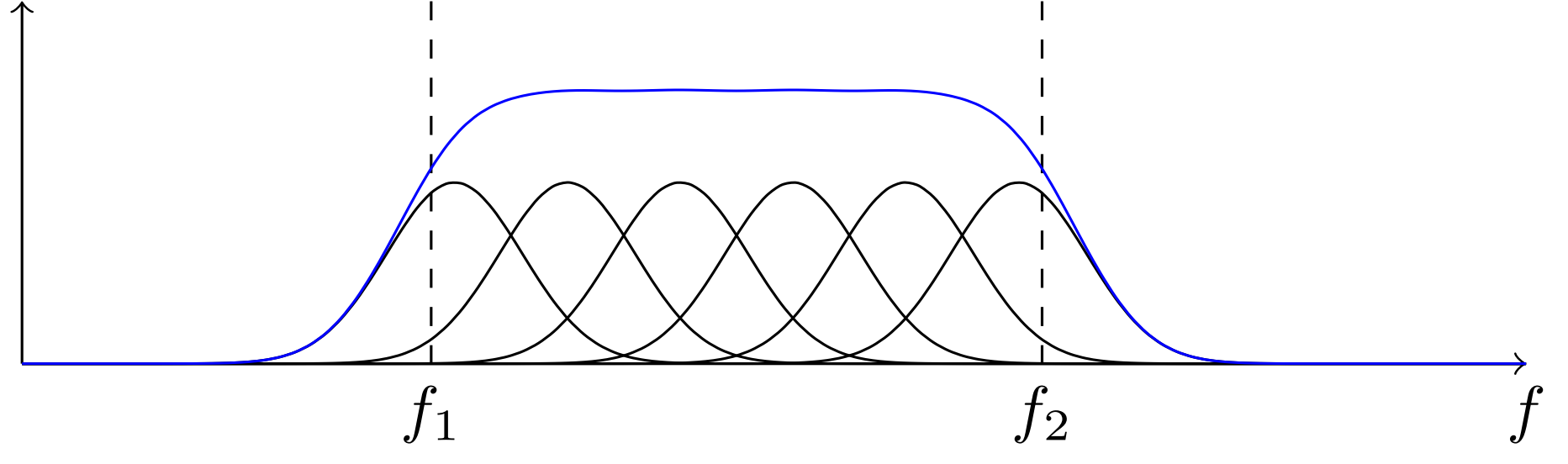}
    \caption{Illustration of a large band Gaussian filter. The top row shows how Gaussian filters can be flattened when the bandwidth is too large. The bottom row illustrates how smaller bandwidth Gaussian filters  can be combined to obtain a better large band filter.}
    \label{fig:largebp}
\end{figure}

Another drawback implied by Gaussian form functions could be the learning of large bandwidths. More precisely, the Gaussian shape is not ideal for large bandwidth filtering as depicted in figure \ref{fig:largebp}. Indeed the frequency response tends to be flat in comparison to a rectangular filter. It is feasible to alleviate this issue by enabling the model to combine smaller band filters (Figure \ref{fig:largebp}). However, and as expected, filters with larger bands are more distributed in the higher frequencies.\\

\section{CONCLUSION}

\textbf{Summary.} This paper introduces and releases a complex-valued and optimal time-frequency resolution alternative to the SincNet architecture. It is based on a complex Gabor filter learning process for automatic speech recognition. The conducted experiments show that the proposed approach is able to produce results comparable with state-of-the-art systems while operating on the raw waveform. 

\noindent\textbf{Future Work.} The base of the Gabor filters in non-orthogonal leading to redundancy in the filtered signals. It is therefore crucial to investigate other filters to produce an orthogonal base for better performances \cite{mallat}. Furthermore, the interest of the phase still has to be established and this model must be evaluated on tasks relying on the phase and local information preserved by complex Gabor filters. 

\section{ACKNOWLEDGEMENTS}
This work was supported by the JST-ANR VoicePersonae Project and by the Engineering and Physical Sciences Research Council (EPSRC) under Grant: MOA (EP/S001530/).

\bibliographystyle{IEEEbib}
\bibliography{strings,refs,bib_titouan}

\end{document}